\documentclass[runningheads,a4paper]{llncs}

\usepackage{amssymb}
\usepackage{graphicx}

\usepackage{tabularx}
\usepackage{booktabs}
\usepackage{cite}
\usepackage{enumitem}
\usepackage[misc]{ifsym}
\usepackage{hyperref}
\usepackage{color}
\usepackage{tikz}
\usepackage{tikzscale}
\usetikzlibrary{positioning,shapes,shadows,arrows}
\usepackage{amsmath}
\usepackage{subcaption}
\usepackage{siunitx}
\usepackage{censor}

\urldef{\mailsa}\path|{t.adler, l.maier-hein}@dkfz.de|

\newif\ifanon
\anonfalse 
\ifanon
\else
  \renewcommand{\xblackout}{}
  \renewcommand{\censor}{}
\fi

\newcommand{\corrAuthor}{$^{(\scriptsize\textrm{\Letter})}$}

\newcommand{\waic}{\operatorname{WAIC}}

\definecolor{myGray}{RGB}{80,80,80}
\definecolor{myYellowAlt}{RGB}{100,100,100}
\definecolor{myGreenAlt}{RGB}{200,200,200}

\begin{document}

\mainmatter  

\title{Out of distribution detection \\for intra-operative functional imaging}

\titlerunning{OoD detection for intra-operative functional imaging }

%
%
\author{Tim J.\ Adler\inst{1,2}\corrAuthor \and Leonardo Ayala\inst{1} \and Lynton Ardizzone\inst{3} \and Hannes G.\ Kenngott\inst{4} \and Anant Vemuri\inst{1} \and Beat P.\ M{\"u}ller-Stich\inst{4} \and Carsten Rother\inst{3} \and Ullrich K{\"o}the\inst{3} \and Lena Maier-Hein\inst{1}\corrAuthor}
\authorrunning{\protect\censor{T.\,J.\ Adler et al.}}



\institute{\xblackout{Division Computer Assisted Medical Interventions (CAMI), German Cancer Research Center (DKFZ), Heidelberg, DE}
\\
\xblackout{\mailsa}
\\
\and \xblackout{Faculty of Mathematics and Computer Science, Heidelberg University, DE}
\\
\and \xblackout{Visual Learning Lab, Heidelberg University, DE}
\\
\and \xblackout{Division of Minimally-invasive Surgery of the Department of General Surgery, Heidelberg University, DE}
}

%
%

\toctitle{OoD detection for intraoperative optical imaging modalities}
\tocauthor{\protect\censor{T.\,J.\ Adler}}
\maketitle
\begin{abstract}
Multispectral optical imaging is becoming a key tool in the
operating room. Recent research has shown that machine learning algorithms can be used to convert pixel-wise reflectance measurements to
tissue parameters, such as oxygenation. However, the accuracy of these
algorithms can only be guaranteed if the spectra acquired during surgery
match the ones seen during training. It is therefore of great interest to
detect so-called \emph{out of distribution} (OoD) spectra to prevent the algorithm from presenting spurious results. In this paper we present an information theory based approach to OoD detection based on the \emph{widely applicable information criterion} (WAIC). Our work builds upon recent methodology related to
\emph{invertible neural networks} (INN). Specifically, we make use of
an ensemble of INNs as we need their tractable Jacobians in order to
compute the WAIC. Comprehensive experiments with \emph{in silico}, and \emph{in vivo} multispectral imaging data indicate that our approach is
well-suited for OoD detection. Our method could thus be an important step towards reliable functional imaging in the operating room.
\end{abstract}
\setcounter{footnote}{0}
\vspace{-1cm}
\section{Introduction}
\label{sec:intro}

The most commonly applied approach to computer aided surgery (CAS) relies on fusing pre-operative medical images with the current patient anatomy for augmented reality guidance. While this approach is well-suited for displaying subsurface structures detected in pre-operative images, such as tumors or vessels, a main bottleneck is the fact that it cannot account for tissue dynamics; live monitoring of perfusion, for example, is not possible with an approach that relies on `offline images'. To address this shortcoming, recent research has focused on intra-operative functional imaging using biophotonics techniques. In this context, multispectral optical imaging is evolving as a key tool. Previous work has shown that machine learning algorithms can be used to convert pixel-wise reflectance measurements to tissue parameters, such as oxygenation~\cite{wirkert_robust_2016,wirkert_physiological_2017}. 
These methods learn to infer tissue parameters via training samples providing a spectrum and the correct corresponding tissue parameter(s) (supervised learning). However, the accuracy of these algorithms is heavily effected by aleatoric and epistimic uncertainties~\cite{kendall2017}.

\begin{figure}[htbp]
    \centering
    \includegraphics[width=0.8\textwidth]{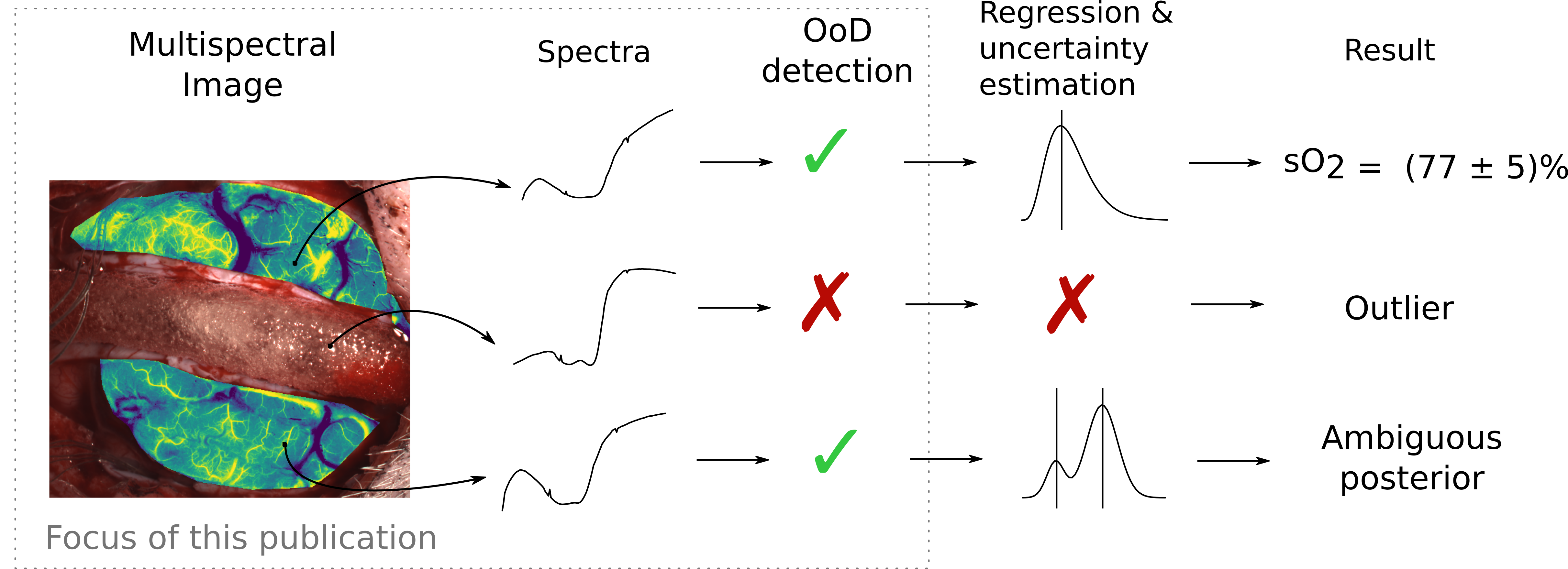}
    \caption{Proposed multi-stage process for uncertainty handling in multispectral image analysis. To investigate whether the input (here: a spectrum) is sufficiently close to the training data, \emph{out of distribution} (OoD) detection is performed. If the input is regarded as valid, the corresponding functional tissue parameters are computed. To address the potential inherent ambiguity of the problem a full posterior probability distribution rather than a point estimate is provided for each tissue parameter (here: blood oxygenation). }
    \label{fig:flow-diagram}
\end{figure}

In this paper, we argue for a multi-stage process for uncertainty handling as illustrated in Fig.~\ref{fig:flow-diagram}. (1) To investigate whether the input is sufficiently close to the training data, \emph{out of distribution} (OoD) detection is performed. (2) If the input is regarded as valid, the corresponding functional tissue parameters are computed, and a full posterior probability distribution is provided as output for each tissue parameter. As the second part of this pipeline has already been addressed in a recent publication~\cite{ardizzone_analyzing_2018, adler2019uncertainty}, we will focus on the first part. The following sections present and validate our proposed approach to OoD detection. 

\vspace{-1.2cm}
\section{Methods}
\label{sec:methods}

While we are not aware of any previous work in OoD detection in the field of optical imaging, the topic has gained increasing interest in the machine learning community. To implement the proposed multi-stage process for uncertainty handling in multispectral image analysis (Fig.~\ref{fig:flow-diagram}), we build our method upon the work by Choi et al.~\cite{choi18_waic_but_why} who proposed the \emph{widely applicable information criterion} (WAIC) as a means to measure the closeness of a new sample to the training distribution. The advantage of this method lies in the fact that it outperforms many other ensemble based unsupervised learning methods while still being easily computable. An unsupervised approach is integral to the method as it is not feasible to generate enough labeled negative samples to train a discriminator between in- and outliers~\cite{choi18_waic_but_why,markou2003novelty}. The challenge in applying WAIC is the fact that it is an ensemble based method leading to the necessity of training a model multiple times. Depending on the data dimensions this can become prohibitively expensive both in terms of time and hardware requirements. In this work, we use  invertible neural networks (INN) \cite{ardizzone_analyzing_2018} to estimate WAIC on multispectral endoscopic imaging data. 

In this section, we briefly revisit WAIC~\cite{choi18_waic_but_why} and give an intuition for this quantity (Section~\ref{ssec:waic}), present the INN architecture as an integral ingredient to apply WAIC in the surgical domain (Section~\ref{ssec:inn}) and  describe our experimental validation (Section~\ref{ssec:setup}).
\vspace{-0.5cm}
\subsection{Principle of WAIC}
\label{ssec:waic}
In the original contribution~\cite{watanabe2009algebraic}, 
WAIC was defined as
\begin{equation}
\waic(x) = \operatorname{Var}_{\Theta}[\log p(x\mid \Theta)] - \mathbb{E}_{\Theta}[\log p(x\mid\Theta)],
\label{eq:waic}
\end{equation}
where \(\waic(x)\)  quantifies the proximity of a sample $x$ to the distribution of the training data \(X^{\text{tr}}\), and \(\Theta\) is distributed according to \(p(\Theta \mid X^{\text{tr}})\) . In a very recent publication~\cite{choi18_waic_but_why} it was suggested to use WAIC as a means for OoD in the setting of neural networks.\footnote{Please note that the sign convention of WAIC of Choi and Watanabe are opposite. We chose Watanabe's definition.} 
The variance term in equation~\eqref{eq:waic} measures `how certain' the posterior distribution \(p(\cdot \mid \Theta)\) is about a sample \(x\), the heuristic being that it should be more certain about samples that are close to what it has seen before. The expectation term in equation~\eqref{eq:waic} is used for normalization. The idea is that if the expectation of \(\log p(x \mid \Theta)\) is high then the spread measured by the variance might also be larger without really measuring internal uncertainty of the model. Hence, it is subtracted to account for this effect. 
\vspace{-.5cm}
\subsection{WAIC computation with INNs}
\label{ssec:inn}
WAIC only works for parametrized models. To meet this precondition, we use a deep neural network to encode the spectra \(X\) in a latent space \(Z\) following an analytically tractable distribution, which we chose to be a multivariate standard Gaussian. Let \(f_\Theta\colon X \subset \mathbb{R}^n \to Z \subset \mathbb{R}^n\) denote the the neural network with parameters \(\Theta\). Then we can use the change of variable formula to compute the log-likelihood \(\log p(x\mid\Theta)\) for a spectrum \(x\) as
\begin{equation}
\log p(x \mid \Theta) = -\frac12 \|f_\Theta(x)\|^2_2 - \frac{n}{2}\log(2\pi) + \log|\det Jf_\Theta(x)|,
\label{eq:ml}
\end{equation}
where \(Jf_\Theta\) denotes its Jacobian~\cite{walter1987real}. Equation~\eqref{eq:ml} shows that it is mandatory for the log-Jacobi determinant of our network to be efficiently computable. One established architecture is the one of \emph{normalizing flows} originally introduced in~\cite{dinh16_densit_estim_using_real_nvp} and refined in~\cite{ardizzone_analyzing_2018} under the name of \emph{invertible neural networks} (INN).
For each of the experiments described in the next section, we trained an ensemble of INNs to estimate \(p(\Theta \mid X^\text{tr})\). Each network consisted of 10 layers of so called coupling blocks (see~\cite{dinh16_densit_estim_using_real_nvp}) each followed by a permutation layer. Each coupling block consisted of a 3 layer fully connected network with ReLU activation functions. The networks were trained using Maximum-Likelihood training, i.\,e.\ by minimizing the loss \(L(x) = -\log p(x \mid \Theta)\) as given in Equation~\eqref{eq:ml} using the Adam optimizer~\cite{kingma2014adam}.

\begin{figure}
    \centering
    \begin{subfigure}[b]{0.6\textwidth}
        \includegraphics[width=\textwidth]{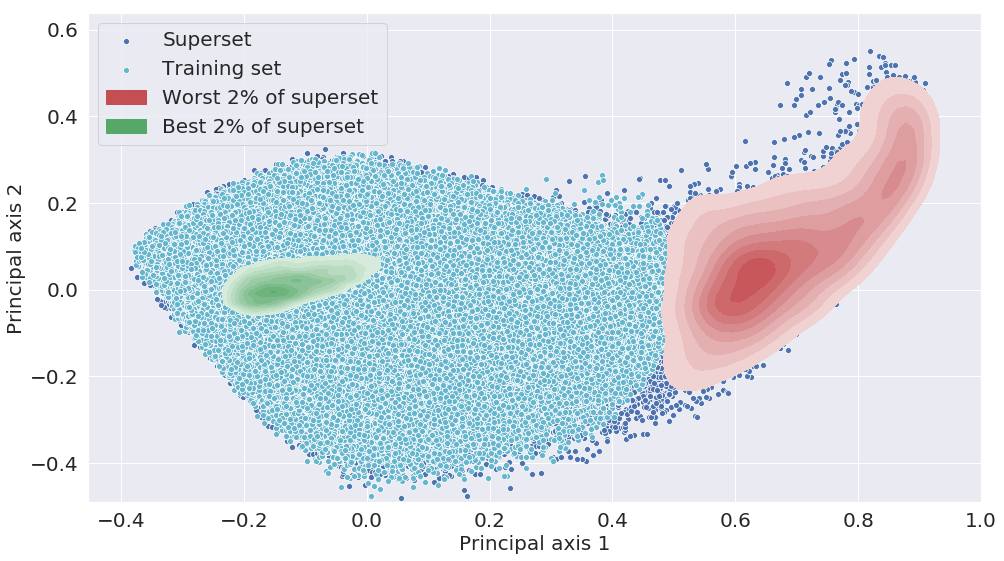}
        \caption{}
    \end{subfigure}
     \begin{subfigure}[b]{0.35\textwidth}
    \includegraphics[width=\textwidth]{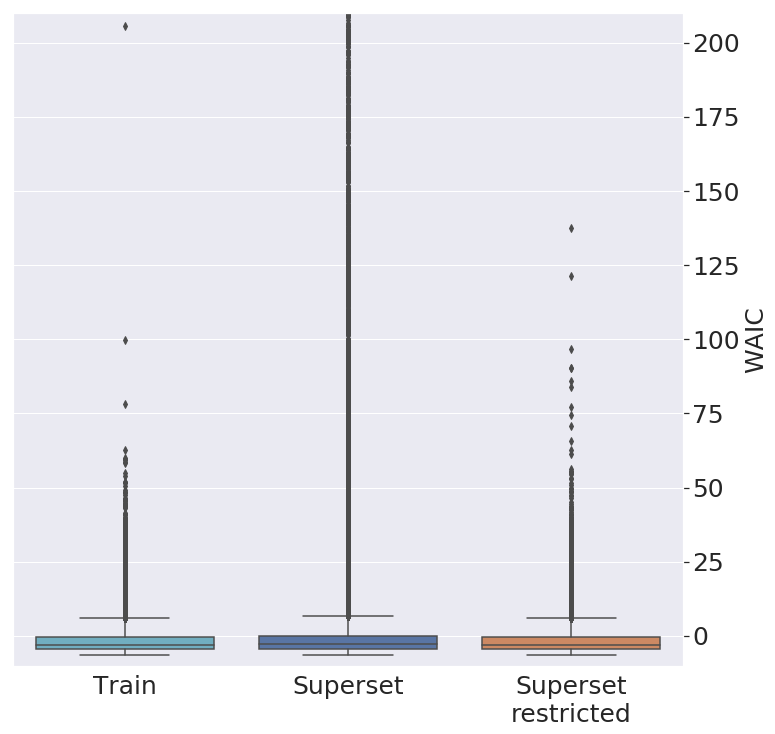}
        \caption{}
    \end{subfigure}
    \caption{In silico validation. (a) The WAIC method trained on the training set (turquoise points, here projected onto the first two principal components (PCA) of the complete training set \(X_\text{SC}^\text{tr}\)) was applied to the superset (here: blue points). The 2
    \% percentile of best and worst superset spectra according the WAIC value are shown as a kernel density estimation in green (representing \textit{in distribution} samples) and red (representing \textit{out of distribution} samples) respectively. (b) The WAIC distribution of the training set, superset and restricted superset is shown (superset boxplot truncated). 
    }
    \label{fig:in-silic-aggregate}
\end{figure}
\subsection{Experiments}
\label{ssec:setup}

The purpose of our experiments was to validate our approach to OoD detection \textit{in silico} (Section~\ref{ssec:in-silico}), and to present \textit{in vivo} use cases (Section~\ref{ssec:application}). 


\subsubsection{In silico quantitative validation}
\label{ssec:in-silico}
In our simulation framework, a multispectral imaging pixel is generated from a 8-valued vector $\mathbf{t}_i$ of tissue properties, which are assumed to be relevant for the image formation process. Plausible tissue samples $\mathbf{t}_i$, are drawn from a layered tissue model as proposed in~\cite{wirkert_physiological_2017}. The framework was used to generate a data set \(X_\text{raw}\), consisting of 550,000 high resolution spectra and corresponding ground truth tissue properties. It was split in a training \(X_\text{raw}^\text{tr}\) and test set \(X_\text{raw}^\text{te}\), comprising 500,000 and 50,000 spectra respectively. For the \emph{in silico} quantitative validation we converted the (high resolution) spectra of the simulated data sets to plausible camera measurements using the filter response functions of the 8-band Pixelteq SpectroCam. We use a subscript (here: \emph{SC} for SpectroCam) to refer to the data set \(X_\text{raw}\) after it was adapted to a certain camera. \(X_\text{SC}^\text{tr}\) was split into a small training set \(X_\text{SC}^\text{tr,s}\) and a \emph{superset} \(X_\text{SC}^\text{sup}\), such that the support of \(X_\text{SC}^\text{tr,s}\) lay within the support of \(X_\text{SC}^\text{sup}\) and \(X_\text{SC}^\text{sup}\) consisted of a cluster of data points outside of the support of \(X_\text{SC}^\text{tr,s}\), as illustrated in Fig.~\ref{fig:in-silic-aggregate} (a). This led to a split of \(X_\text{SC}^\text{tr,d}\) of approximately 49\% of \(X_\text{SC}^\text{tr}\) and \(X_\text{SC}^\text{sup}\) of approximately 51\% of \(X_\text{SC}^\text{tr}\). An ensemble of five INNs was trained on \(X_\text{SC}^\text{tr,s}\) and the WAIC value was evaluated on \(X_\text{SC}^\text{sup}\). We defined \(X_\text{SC}^\text{sup,r}\) as the \emph{reduced} data set of \(X_\text{SC}^\text{sup}\) lying in the support of \(X_\text{SC}^\text{tr,s}\). We then investigated (1) whether the WAIC distribution of the \(X_\text{SC}^\text{sup,r}\) matches that of the \(X_\text{SC}^\text{tr,s}\) and whether (2) the part of \(X_\text{SC}^\text{sup}\) not in the support of \(X_\text{SC}^\text{tr,s}\) was correctly classified as outliers by our method.

\vspace{-.5cm}
\subsubsection{In vivo application}
\label{ssec:application}

While the goal of the previous experiment was to confirm the validity of our approach in an \textit{in silico} setting, the purpose of the \textit{in vivo} experiments were to showcase applications in which the OoD detection could be useful.

\textbf{Anomaly/Novelty detection:} Detecting (parts of) a multispectral image in which the spectra do not closely match the training data distribution can be useful for many reasons. Possible applications include the detection of abnormal tissue or of artifical objects (e.\,g.\ instruments). To investigate this aspect, we used the complete \(X_\text{SC}^\text{tr}\) to train an ensemble of five INNs. As \textit{in vivo} test data, we acquired endoscopic images of porcine organs which we classified as \emph{organs lying in the simulation domain} \(X^\text{iD}\) and \emph{organs not lying in the simulation domain} \(X^\text{oD}\). These spectra were acquired using a Pixelteq SpectroCam on a \SI{30}{\degree} Stortz laparascope with a a Stortz Xenon light source (Storz D-light P 201337 20). We classified liver, spleen, abdominal wall, diaphragm and bowl as in domain organs as  hemoglobin can be assumed to be the main absorber in these. In contrast, we classified gallbladder as an out of domain organ, since bile is a notable absorber but has not been considered in our simulation framework. With this, \(X^\text{iD}\) and \(X^\text{oD}\) consisted of 50000 spectra and 10000 spectra respectively. Our hypothesis was that the WAIC values of \(X^\text{iD}\) should be much lower than those for \(X^\text{oD}\). For reference, we also compared the resulting WAIC distributions to that of the simulated test data \(X_\text{SC}^\text{te}\).

\begin{figure}[htbp]
    \centering
    \includegraphics[width=.9\textwidth]{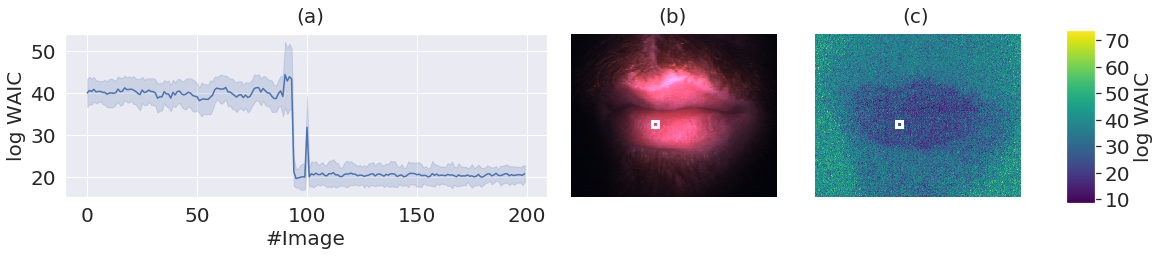}
    \caption{Automatic detection of scene changes. (a) When a change in light source occurs, the mean log WAIC values computed for the white region of interest in (b)/(c) drop, indicating a decreasing domain gap between training and test data. (b) RGB image estimated using the 8-band measurement of human lips. (c) Corresponding WAIC values computed for the multispectral image.}
    \label{fig:lip_map}
    \vspace{-.5cm}
\end{figure}

\textbf{Detection of scene changes:} Intra-operative image modalities often rely on a careful calibration of the device. When recovering blood oxygenation from multispectral measurements, for example, the regressor is typically trained with the light source that is used during test time. To investigate whether WAIC is applicable to detect illumination changes (which would substantially harm the method and render the estimation results invalid), we adapted \(X_\text{raw}\) to a xiQ XIMEA (Muenster, Germany) \(SNm4\times4\) mosaic camera consisting of 16 bands assuming a Wolf LED light source (Wolf Endolight LED 2.2). We trained an ensemble of five INNs on \(X_\text{Xim}^\text{tr}\). Furthermore, we recorded 200 \(512\times272\)-pixel images of the lip of a healthy human volunteer using the xiQ XIMEA camera and a \SI{30}{\degree} Stortz laparascope (cf.\ Fig.~\ref{fig:lip_map} (b)). At around image 80 we switched the endoscope from a Stortz Xenon light source (Storz D-light P 201337 20) to a Wolf LED light source (Wolf Endolight LED 2.2). Based on the hypothesis that the switch in light source would be detected by WAIC analysis, we computed the WAIC time series for the region of interest depicted in Fig.~\ref{fig:lip_map} (a).

\section{Results}

\begin{figure}[htbp]
    \centering
    \includegraphics[width=0.8\textwidth]{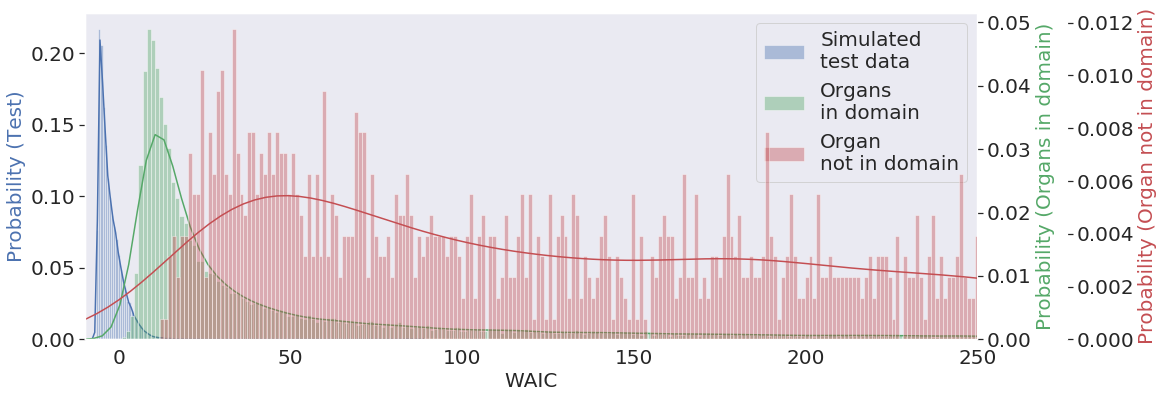}
    \caption{Histogram of the WAIC values for the simulated test set, and \textit{in vivo} multispectral measurements of organs that do (green) or do not (red) match the model assumptions based on which the training data was generated (tails truncated). Please note the different scales for the three distributions.}
    \label{fig:organ_waic_histogram}
\end{figure}
\vspace{-.2cm}
\textbf{In silico validation} The distribution of the reduced training data \(X_\text{SC}^\text{tr,s}\) and the superset \(X_\text{SC}^\text{sup}\) (projected to the first two principal components) can be found in Fig.~\ref{fig:in-silic-aggregate}~(a). It can be seen that the samples with poor WAIC values (red) concentrate in the superset part not contained in \(X_\text{SC}^\text{tr,s}\), whereas samples with low WAIC values (green) are contained in the interior of \(X_\text{SC}^\text{tr,s}\). Fig.~\ref{fig:in-silic-aggregate}~(b) shows a comparison between the WAIC distributions of \(X_\text{SC}^\text{tr,s}\), \(X_\text{SC}^\text{sup}\) and the restricted superset \(X_\text{SC}^\text{sup,r}\). The data sets \(X_\text{SC}^\text{tr,s}\) and \(X_\text{SC}^\text{sup,r}\) are in excellent agreement. The superset \(X_\text{SC}^\text{sup}\) only differs in the regard that there are far more outliers, which can be accounted to the data points outside of \(X_\text{SC}^\text{tr,s}\).

\textbf{Application} The WAIC distribution for the test data \(X_\text{SC}^\text{te}\), the \emph{in domain organs} \(X^\text{iD}\) and the \emph{out of domain organ} \(X^\text{oD}\) can be found in Fig.~\ref{fig:organ_waic_histogram}. The distribution of the test data is by far the sharpest with a maximum aposterior probability (MAP) at -4.9. The distribution closely matches that of the training data (MAP = -4.9). The distribution of \(X^\text{iD}\) also possesses a sharp maximum, however with a far heavier tail. The MAP estimate yields 9.3 which indicates a still existent domain gap between our simulation domain and the organ domain. The distribution of \(X^\text{oD}\) is very noisy and has an even heavier tail than \(X^\text{iD}\). The MAP lies at 34. This indicates that our WAIC estimate is suitable to distinguish between in domain tissue and out of domain tissue. Similarly, Fig.~\ref{fig:lip_map} illustrates that the change in illumination as performed in the \textit{detection of scence changes} experiment results in a drastic change of WAIC values. 
\vspace{-0.2cm}
\section{Discussion}
\label{sec:discussion}

The accuracy of machine learning-based regression methods in multispectral imaging crucially depend on whether the spectra acquired during surgery match the ones seen during training. Although initial steps with respect to uncertainty estimation and compensation have been taken in the field of optical imaging~\cite{ardizzone_analyzing_2018,adler2019uncertainty, kohl_probabilistic_2018, zhu_bayesian_2018, leibig_leveraging_2017, gal_dropout_2016}, we are, to our knowledge, the first to address the problem of OoD detection to prevent algorithms from presenting spurious results.  

The application to endoscopic organ data showed that our method is well-suited for anomaly detection. The in distribution organs are well separated from the out of distribution organ (gallbladder). Moreover, this experiment reveals a shortcoming of the simulation framework proposed by~\cite{wirkert_physiological_2017}: The large difference in the WAIC distribution between the test set and the real data indicates a domain gap that remains to be tackled.

Our experiments with human lips show that WAIC is able to distinguish between different lighting conditions. The uncertainty prior to the change of lighting can most likely be explained by the short darkness stemming from the light source switch. The jump at image 100 was due to involuntary movement of the volunteer leading to the image being out of focus. 
 One reason for the generally high WAIC values is the fact that melanin (a chromophore in the skin) was not simulated in the training data.

In the present implementation we used five INNs in our ensembles. According to preliminary experiments, this number is sufficient. We computed the WAIC on the data sets used for the anomality detection experiment (Section~\ref{ssec:application}) for up to 20 ensemble members. For both the simulated test data and the in domain organs the values stabilized below \(n = 10\). For the out of domain data (gallblader) the WAIC values increased throughout. This merits further investigation. However, there should be no impact on the method performance, as \(X^\text{iD}\) and \(X^\text{oD}\) were well separated.

Our findings underline the power of WAIC in the setting of medical OoD detection. However, there are still some open questions. A general downside of WAIC is its `arbitrary units' and it is not straightforward to define a threshold for outlier detection. One approach to tackle this shortcoming would be to find a suitable normalization. Another possibility might be to just mask the worst \(n\) pixels in a certain ROI. 
Additionally, to this conceptual question, there are also practical limitations. The estimation of WAIC requires an \emph{ensemble} of neural networks, which was feasible in our case, but becomes prohibitively expensive for larger input dimensions. For the future, methods for network compression might be adapted to tackle this problem.

In conclusion, this paper is the first to address the topic of OoD detection in intra-operative imaging. Due to the promising results obtained in this study, the approach proposed could not only become a valuable tool for increasing the reliability of machine learning-based regression methods but could also boost research in unsupervised intra-operative anomaly detection.

\vspace{-.4cm}
\ifanon
\bibliography{main_short_anon}
\else
\bibliography{main_short}
\fi
\end{document}